\documentstyle[aps,floats,eqsecnum,amsfonts,epsfig,times]{revtex}

\newcommand{\lapp}{\mathrel{\vcenter{\hbox{\tiny \ooalign{\raise 3.25pt
        \hbox{$<$}\crcr $\sim$}}}}}
\newcommand{\gapp}{\mathrel{\vcenter{\hbox{\tiny \ooalign{\raise 3.25pt
        \hbox{$>$}\crcr $\sim$}}}}}
\newcommand{\eqdef}{\!\!\mathrel{\vcenter{\hbox{ \ooalign{\raise 4.75pt
        \hbox{${\textsf{\tiny{\,def}}}$}\crcr $=$}}}}}

\newcommand{\bi}{\begin{itemize}}
\newcommand{\ei}{\end{itemize}}

\newcommand{\forget}[1]{\iffalse#1\fi}
\newcommand{\forgetmenot}[1]{\iftrue#1\fi}

\newcommand{\be}{\begin{equation}}
\newcommand{\ee}{\end{equation}}

\renewcommand{\:}[2]{{\textstyle\frac{#1}{#2}}}
\renewcommand{\;}[2]{{\frac{#1}{#2}}}
\newcommand{\ba}{\begin{eqnarray}}
\newcommand{\ea}{\end{eqnarray}}
\newcommand{\li}{&}
\newcommand{\bra}[1]{\left(#1\right)}
\newcommand{\bras}[1]{\left[#1\right]}
\newcommand{\brac}[1]{\left\{#1\right\}}

\def\d{\dot}
\def\dd{\ddot}

\title{Qualitative Properties of Magnetic Fields in Scalar Field
Cosmology}

\author{C.A. Clarkson, A.A. Coley and S.D. Quinlan.\footnote{Email:
clarkson@mathstat.dal.ca, aac@mathstat.dal.ca, quinlan@mathstat.dal.ca.}}
\address{Department of Mathematics and Statistics, Dalhousie
University, Halifax, NS, Canada, B3H 3J5.}
\date{\today}

\begin{document}
\maketitle

\begin{abstract}
We study the qualitative properties of the class of spatially homogeneous
Bianchi  VI$_{\mbox{o}}$ cosmological models containing a perfect fluid with a
linear equation of state, a scalar field with an exponential potential and a
uniform cosmic magnetic field, using dynamical systems techniques. We find that
all models evolve away from an expanding massless scalar field model in which
the matter and the magnetic field are negligible dynamically. We also find that
for a particular range of parameter values the models evolve towards the usual
power-law inflationary model (with no magnetic field) and, furthermore, we
conclude that inflation is not fundamentally affected by the presence of a
uniform primordial magnetic field.  We investigate the physical properties of
the Bianchi~I magnetic field models in some detail.

\end{abstract}

\section{Introduction}

There are many observations that imply the existence of magnetic fields in the
Universe, and recent observations have produced firmer estimates of the
strength of magnetic fields in interstellar and intergalactic space (Kronberg,
1994; Kronberg et al., 1992; Wolf et al., 1992). The observations to date place
an upper bound on the strength of a cosmic magnetic field, but are not
conclusive as to whether such a field exists (see Vall\'ee 1990a,b). Although
cosmologists have investigated the possible existence of a homogeneous
intergalactic magnetic field of primordial origin both from a theoretical and
an observational point of view for many years, research on cosmological
magnetic fields has been rather marginal despite their potential importance.
The reasons could be the perceived weakness of the field effects or the lack,
as yet, of a consistent theory explaining the origin of cosmic magnetism.
However, this situation has changed considerably recently (cf. Grasso and
Rubenstein, 2000).

Primordial magnetic fields introduce new ingredients into the standard picture
of the early Universe. Such a field would affect  the temperature distribution
of the microwave background radiation, primeval nucleosynthesis and galaxy
formation. But its most direct observational effect is the Faraday rotation it
would cause in linearly polarized radiation from observed extragalatic radio
sources. Fundamental properties of magnetic fields include their vectorial
nature, which inevitably couples the field to the spacetime geometry, and the
resulting tension (i.e., the negative pressure) exerted along the field's lines
of force (Parker, 1979; Matravers and Tsagas, 2000 ). The implications of such
an interaction are both kinematical and dynamical; kinematically, the
magneto-curvature effect tends to accelerate positively curved perturbed
regions, while it decelerates regions with negative local curvature and,
dynamically, the most important magneto-curvature effect is that it can reverse
the pure magnetic effect on density perturbations (Tsagas and Barrow, 1997 and
1998; Tsagas and Maartens, 2000 and 2001).

In a recent analysis (Matravers and Tsagas, 2000) the kinematics were
considered and, assuming a spacetime filled with a perfectly conducting
barotropic fluid permeated by a weak primordial magnetic field while treating
the energy density and the anisotropic pressure of the field as first-order
perturbations upon the Friedmann-Lema\^\i tre-Robertson-Walker (FLRW)
background, it was shown how a cosmological magnetic field can modify the
expansion rate of an almost-FLRW universe. The effects of the interaction
between magnetism and geometry in cosmology are subtle, and it was argued that
the magneto-curvature coupling could make the field into a key player
irrespective of the magnetic strength. Indeed, even weak magnetic fields can
lead to appreciable effects, provided that there is a strong curvature
contribution. This was illustrated by studying spatially open cosmological
models containing `matter' with negative pressure, and it was found that the
phase of accelerated expansion, which otherwise would have been inevitable, may
not even happen. This leads to the question of the {\it efficiency of
inflationary models in the presence of primordial magnetism}. Recall that an
initial curvature era was never considered as a problem for inflation, given
the smoothing power of the accelerated expansion. However, this may not be the
case when a magnetic field is present, no matter how weak the latter is. In
more recent work Tsagas (2000) has further studied the vector nature of
magnetic fields and discussed the unique coupling between magnetism and
spacetime curvature in general relativity which gives rise to a variety of
effects with important implications.

Observations of the high degree of isotropy of the cosmic microwave background
indicate that the Universe is `almost' isotropic and spatially homogeneous (at
least since the time of last scattering). Theoretical support for this belief
comes from the so-called Ehlers-Geren-Sachs (1964) Theorem. Clarkson and Coley
(2001) have proven a generalization of this theorem and have consequently shown
that any strong magnetic fields in the Universe are ruled out. This theoretical
result is model-independent and includes the case of inhomogeneous magnetic
fields. In further work (Clarkson et al., 2001) numerical constraints are
placed on all types of primordial and protogalactic magnetic fields in the
Universe from cosmic microwave background data.

We shall study open spatially homogeneous cosmological models containing both a
uniform magnetic field and a scalar field here, partially in an attempt to
address some of the questions raised above. Scalar field cosmological models
are of great importance in the study of the early Universe.  Models with a
variety of self-interaction potentials have been studied, and one potential
that is commonly investigated and which arises in a number of physical
situations has an exponential dependence on the scalar field (Wetterich, 1988;
Ferreira and Joyce, 1997). There have been a number of studies of spatially
homogeneous scalar field cosmological models with an exponential potential,
with particular emphasis on the possible existence of inflation in such models
(Halliwell, 1987; Kitada and Maeda, 1993; and Billyard et al., 1999, and
references within).

Dynamical systems methods for analysing the qualitative properties of
cosmological models have proven very useful (Wainwright and Ellis, 1997; Coley,
1999). A qualitative analysis of cosmological models with matter and a uniform
magnetic field has been presented previously (Collins, 1972; LeBlanc et al.,
1995). A universe with a primordial magnetic field is necessarily anisotropic.
Thus, in order to investigate the influence of the magnetic field on the
dynamics of the universe  the Einstein field equations must be analysed in
models more general than the FLRW models. The simple classes of the Bianchi I
and Kantowski-Sachs models were discussed (cf. Jacobs, 1969), and Collins
(1972) applied techniques from the theory of planar dynamical systems to prove
qualitative results about the evolution of the class of axisymmetric Bianchi I
cosmologies with matter and a primordial magnetic field.

In LeBlanc et al. (1995) the Einstein-Maxwell field equations for orthogonal
Bianchi VI$_{\mbox{o}}$ cosmologies with a $\gamma$-law perfect fluid  and a
pure, homogeneous source-free magnetic field were written as an autonomous
differential equation in terms of expansion-normalized variables. It was shown
that the physical region of state space is compact, and that the differential
equation admits certain invariant sets and monotone functions which play an
important role in the analysis. A complete analysis of the stability properties
of the equilibrium points of the differential equation and a description of the
bifurcations that occur as the equation of state parameter $\gamma$ varies was
given . The associated dynamical system was studied and the past, intermediate
and future evolution of these models was determined. All asymptotic states of
the models, and the likelihood that they will occur, were described.  In
particular, oscillatory behaviour also occurs in cosmological models with a
magnetic field (and in Einstein-Yang-Mills theory in general). Further work on
spatially homogeneous models with a magnetic field and a non-tilted perfect
fluid has been carried out recently (LeBlanc, 1997 and 1998). In particular,
Weaver (2000) has generalized (to the non-polarized solutions) the work of
Leblanc et al. (1995), and rigorously shown that the evolution toward the
singularity is oscillatory in Bianchi VI$_0$ vacuum models.

The reasons for concentrating on the Bianchi VI$_{\mbox{o}}$ models are ones
primarily of mathematical simplicity.  It is well known (Hughston and Jacobs
1970) that a pure magnetic field is only possible in Bianchi cosmologies of
types I, II, VI$_{\mbox{o}}$, VII$_{\mbox{o}}$ (in class A) and type III (in
class B). For types VI$_{\mbox{o}}$ and  VII$_{\mbox{o}}$, the algebraic
constraints that arise from the Einstein field equations imply that the shear
eigenframe is Fermi-propagated, which in turn implies that the remaining field
equations reduce to an autonomous differential equation with a polynomial
vector field. Finally, for type VI$_{\mbox{o}}$ but not for VII$_{\mbox{o}}$,
the physical region of state space is compact.  We note that since a magnetic
field is not compatible with Bianchi types VIII and IX (for example), the class
of Bianchi VI$_{\mbox{o}}$ magnetic cosmologies is of the same generality as
the Bianchi VIII/IX models (without magnetic field) (Wainwright 2000).

In this paper we  shall investigate the class of Bianchi  VI$_{\mbox{o}}$
models with barotropic matter, a scalar field with an exponential potential and
a uniform magnetic field. We shall discuss some general properties of the
complete class of models, and then investigate the special case of the Bianchi
I system in detail. We note that this is not the general class of Bianchi I
cosmological models; we are only considering those Bianchi I models that occur
as subcases of the Bianchi VI$_{\mbox{o}}$ models with one magnetic degree of
freedom. It is also of interest to study the Bianchi II cosmologies, since
although they are very special within the full Bianchi class, they play a
central role since the Bianchi II state space is part of the boundary of the
state space for all higher Bianchi types (i.e., all types except for I and V).
The Bianchi II models will be investigated in detail elsewhere (Quinlan, 2002).

The Bianchi I system in the absence of a scalar field was first qualitatively
analysed in (Collins, 1972). A qualitative analysis of the Bianchi I models in
the absence of a magnetic field was given in (Billyard et al., 1999); in this
work the well-known power-law inflationary solution (Wetterich, 1988; Kitada
and Maeda, 1993)  was shown to be a stable attractor for an appropriate
parameter range in the presence of a barotropic fluid in all Bianchi class B
models (previous analysis had shown that this power-law inflationary solution
is a global attractor in spatially homogeneous models in the absence of a
perfect fluid,  except for a subclass of Bianchi type IX models which
recollapse). One of the aims of the present analysis is to study the stability
of this model with respect to magnetic field perturbations.

\section{The Bianchi VI$_{\mbox{o}}$ Models}

We shall follow the approach of  LeBlanc et al. (1995) in which  the theory of
dynamical systems was used to give a detailed analysis of the evolution of
orthogonal Bianchi cosmologies of type VI$_{\mbox{o}}$ with a perfect fluid and
a magnetic field as source.  This work extended that of Wainwright and Hsu
(1989) who had studied perfect fluid Bianchi cosmologies of class A using state
variables that are dimensionless and have a direct physical or geometric
interpretation in terms of the shear of the cosmological fluid, the spatial
curvature and the magnetic field, leading to a state space that is a compact
subset of ${\mathbb{R}}^5$, which implies that a unified treatment of the
asymptotic behaviour of the models at early and late times can be given. We
note that all of the equilibrium points of the differential equation correspond
to self-similar exact solutions of the Einstein field equations.

An invariant orthonormal frame of vector fields on the  spacetime are
introduced in which one vector is aligned along the fluid flow vector so that
the remaining three spatial vectors (triad) span the tangent space orthogonal
to the fluid flow at each point of the group orbits. The commutation functions
of this frame are then taken as the basic variables. The Einstein-Maxwell field
equations for a pure magnetic field are then derived in the orthonormal frame
formalism, giving rise to a set of evolution equations for the shear variables
($\sigma_{\alpha\beta}$), the curvature variables ($n_{\alpha\beta}$), the
energy density (conservation equation), the magnetic field (Maxwell equation)
and a first integral (the Friedmann constraint), where the remaining freedom in
the choice of spatial tetrad was used to simultaneously diagonalize
$\sigma_{\alpha\beta}$ and $n_{\alpha\beta}$.

Expansion-normalized variables are then introduced, leading to a reduced set of
evolution equations (the dynamical system) for the dimensionless shear
variables $\Sigma_{\pm}$, the spatial curvature variables $N_\alpha$, the
dimensionless magnetic field variables and the density parameter (defined as
usual). Defining a dimensionless time variable, the reduced evolution equations
in the special case of a Bianchi VI$_{\mbox{o}}$ were then established (in this
case there is a single dimensionless magnetic field variable, $M$).

\subsection{The Equations}

We assume a scalar field with an exponential potential $V = V_0 e^{k
\phi}$,
where $V_0$ and $k$ are positive constants, and a separately conserved perfect
which satisfies the barotropic equation of state $p = (\gamma -1) \rho$, where
the constant $\gamma$ satisfies $0\le\gamma\le 2$ (although we shall only be
interested in the range $0<\gamma < 2$ here). The energy conservation and
Klein-Gordon equations become:
\ba
\d{\rho}   &    =  &   -3 \gamma H \rho,\\
\dd{\phi}  &   =  &   -3H\d{\phi} - k V,
\ea
where $H$ is the Hubble parameter, an overdot denotes ordinary differentiation
with respect to time $t$, and units have been chosen so that $8 \pi G =1$.

We also assume a uniform magnetic field (in the `$x$' direction) which
satisfies the Maxwell equations. The form of this magnetic field, which we
shall denote here as $M^2$, is given in Collins (1972) or LeBlanc et al.
(1995), but its precise form is unimportant since we use the generalized
Friedmann constraint to eliminate $M^2$ from the evolution equations.

We define
\be
\Omega \equiv \frac{\rho}{3H^2},~~~\Phi \equiv
\frac{\dot{\phi}}{\sqrt{6}H},~~~\Psi \equiv \frac{V}{3H^2},
\ee
two normalized (dimensionless) shear variables $\Sigma_+$ and  $\Sigma_-$
(appropriately combinations of the derivatives of the metric functions divided
by the Hubble parameter; Wainwright and Ellis, 1997)), and the new logarithmic
time variable $\tau$ by
\be
\frac{d \tau}{dt} \equiv H.
\ee

The evolution equations for the quantities
\be
 X= \left( \Sigma_+, \Sigma_-,\Phi, \Psi, \Omega,N_+,N_-\right)\in
{\mathbb{R}}^7,
\ee
are then as follows:
\ba
\Sigma_+' \li = \li (q - 2)\Sigma _{+} + 2(1 - N^2_1  -\Phi ^{2} - \Psi -
\Sigma_+^{2} -
\Sigma_{-}^2 - \Omega)-2N_-^2,\\
\Sigma_-'\li=\li (q - 2)\Sigma_{-}-2N_+N_-,\\
\Phi'\li =\li (q - 2)\Phi  -\frac{\sqrt{6}k}{2}\Psi,  \\
\Psi'    \li =\li (2q+2  + \sqrt{6}k\Phi)\Psi, \\
\Omega' \li  =\li (2q - 3\gamma  + 2)\Omega,\\
 N_+'\li=\li(q+2\Sigma_+)N_++6\Sigma_-N_-,\\
 N_-'\li=\li(q+2\Sigma_+)N_-+2\Sigma_-N_+.
\ea
where a prime denotes differentiation with respect to the time $\tau$ and the
deceleration parameter $q$ is defined by $q\equiv-(1+H'/H)$, where
\be
q  = 1-N_-^2 + \Sigma_+^{2} + \Sigma_{-}^{2} + \frac{1}{2}(3\gamma  - 4)\Omega
+ \Phi ^{2} - 2\Psi.
\ee
The physical state space is restricted to $\Omega \geq 0$ and $N^2_+ \leq
3N^2_-$ (where we include equality to include models of Bianchi types I and II;
LeBlanc et al., 1995). Since $N_-=0$ implies $N^\prime_- = 0$, the subsets $N_-
> 0$ and $N_- < 0$ are invariant. Due to a discrete symmetry in evolution
equations we can, without loss of generality, restrict attention to the subset
$N_- \leq 0$.

The magnetic field, defined through the first integral, is given by
\be
\frac{3}{2} M^2 = 1-N_-^2 - \Phi ^{2} - \Psi - \Sigma_+^{2} - \Sigma_{-}^2
-
\Omega,
\ee
and since $M^2\ge 0$, all of the variables are bounded:
\be
0 \le\left\{ \Sigma_+^{2}, \Sigma_{-}^{2}, \Phi ^{2},\Psi, \Omega, N_+^2/3,
-N_- \right\} \le 1.
\ee
Without loss of generality, we can restrict attention to the subset $M
\geq 0$.
There is also an auxiliary equation for the magnetic field:
\be
M'=(q-1-2\Sigma_+)M.
\ee

\subsection{Discussion}

A number of invariant sets of the physical state space can be identified (cf.,
LeBlanc et al., 1995, and Billyard et al., 1999).  In addition, there are a
number of monotonic functions that exist in various invariant sets
(particularly on the boundary; cf. p.521 LeBlanc et al., 1995).  However, we
shall not present them here. Indeed, there are many (over fifty) equilibrium
points, the vast majority of which are saddles, and hence in this section we
shall simply comment on some of the more physically interesting local
properties of the  models. In the next section we shall investigate the Bianchi
I models more rigorously, and  a monotonic function in this invariant set will
be given explicitly, in order to discuss some aspects of the intermediate or
transient behaviour of the models.

\begin{enumerate}
\item[$\bullet$] The equilibrium point ${\cal P}: (\Sigma_+, \Sigma_-,
\Phi, \Psi, \Omega, N_+, N_ -) = (0,0, \frac{-k}{\sqrt{6}}, 1 -
\frac{k^2}{6},
0, 0,0),~M=0$ corresponds to the power-law inflationary zero-curvature (flat)
FLRW model.  The eigenvalues of ${\cal P}$ are $\{k^2 -4, k^2 -3 \gamma,
\frac{1}{2} (k^2 -2) (\times 2), \frac{1}{2}(k^2 -6)(\times 3)\}$. ${\cal
P}$ exists for $k^2 \leq 6$ and is a sink for $k^2<4$ and  $k^2<3\gamma$, and
is inflationary $(q<0)$ for  $k^2<2$. In fact it is the global attractor for
this range of parameter values; this result generalizes previous work by
including a cosmic magnetic field (cf. Billyard et al., 1999).

\item[$\bullet$] We are particularly interested in whether there are
any sinks (or sources) with $M \neq 0$ in the Bianchi VI$_{\mbox{o}}$ state
space. We know from LeBlanc et al. (1995) that there are no attractors in the
absence of a scalar field, so hence we must have $M \neq 0$, $\Phi^2 \neq 0$
(and $\Psi \neq 0$, since this leads to $M =0$).  From equation (2.10) we see
that $\Omega$ approaches zero at late times, so we also take $\Omega =0$ (and
$q < 2$).  From equations (2.8), (2.9) and (2.16) we have that (for $M
\neq 0$)
\begin{equation}
\Phi = \frac{-4}{\sqrt{6}k} (1+ \Sigma_+),~~~\Psi =
\frac{4}{3k^2}(1+\Sigma_+)
(1-2\Sigma_+); \quad q = 1 + 2 \Sigma_+.
\end{equation}
We are particularly interested in the inflationary case in which $q < 0$; i.e.,
$-1 < \Sigma_+ < -\frac{1}{2}$. If $N_-, N_+ \enskip (\Sigma_-)$ are not zero,
equations (2.11) and (2.12) yield $N_+ = \pm \sqrt{3}N_-$, $\Sigma_- = \mp
\frac{1}{2\sqrt{3}} (1+4 \Sigma_+)$, whence equation (2.7) then yields
$N^2_- =
\frac{1}{12} (1-2 \Sigma_+) (1+4\Sigma_+)$, which is negative for the
range of $\Sigma_+$ under consideration! (There do exist equilibrium points
with $N_-
\ne 0, N_+\ne 0$ and $\Sigma_-\ne 0$, but these all have $q \ge 0$ (i.e.,
are non-inflationary) and are {\it saddles}.) Hence we consider $N_- = N_+ =
\Sigma_-=0$. Equations (2.6) and (2.13) then yield (since $\Sigma_+ \neq
\frac{1}{2}$), $\Sigma_+ =-1 + 3k^2/(8+k^2)$. This equilibrium point,
denoted ${\cal V}$ in section V, is physical and in the state space when $4 <
k^2 <8$ and is always a saddle $(N^2_+$ is increasing as orbits evolve away
from ${\cal V}$ to the future). However, since $q < 0$ only for $k^2 <
\frac{8}{5}~(\Sigma_+ < -\frac{1}{2})$, this point is never inflationary.
\end{enumerate}

We conclude that inflation is not fundamentally affected by the presence of
primordial magnetism in the models under study here.

\begin{enumerate}
\item[$\bullet$] In LeBlanc et al. (1995) it was shown that in the
absence of a scalar field that for $\Omega >0$, $1 < \gamma <2$ the global sink
is the equilibrium point $PM$(VI$_{\mbox{o}}$) so that at late times all such
magnetic Bianchi VI$_{\mbox{o}}$ cosmologies are approximated by the dynamics
of a self-similar Dunn-Tupper (1980) magneto-vacuum Bianchi VI$_{\mbox{o}}$
model.  This equilibrium point has an analogue here with $\Phi = \Psi =0$, but
is now a saddle.  Indeed, it can be easily shown that none of the equilibrium
points with vanishing scalar field can be sinks.  In particular, the sink in
the case $\gamma =1$ in LeBlanc et al. (1995) (subset of ${\cal
L}M$(VI$_{\mbox{o}}$)) becomes a saddle.

\item[$\bullet$]  All equilibrium points corresponding to models with a
non-zero matter fluid, including those corresponding to the flat FLRW perfect
fluid solution (Einstein de Sitter) and the flat FLRW scaling solution, are
saddles.
\end{enumerate}

In the above we have made some passing comments concerning the intermediate
behaviour of the models, but we have not discussed this in detail. In the next
section we will discuss the intermediate behaviour of the (subset) of Bianchi I
models. The transient behaviour is important for assessing the physical
significance of the models. For example,  LeBlanc et al. (1995) also considered
whether any of the Bianchi VI$_{\mbox{o}}$ magnetic cosmologies are compatible
with observations of a highly isotropic universe. For a subset of the Bianchi
VI$_{\mbox{o}}$ cosmologies, the orbits will initially be attracted to the
equilibrium point ${\cal F}$ corresponding to the flat FLRW model, but will
eventually be repelled by it.  For such a model there will be an epoch during
which the model is close to isotropy and hence will be compatible with
observations.  This is the phenomenon of {\it intermediate isotropization}.  It
will not occur for a randomly selected magnetic Bianchi VI$_{\mbox{o}}$ model,
but there will be a finite probability that an arbitrarily selected model will
undergo intermediate isotropization (the probability will depend on the desired
closeness to isotropy and the length of the epoch of isotropization).

\begin{enumerate}
\item[$\bullet$]
There is an equilibrium set ${\cal K}$: $ (\Sigma_+, \Sigma_-, \Phi, 0,0,0,0)$
where $1-\Sigma^2_+ - \Sigma^2_- - \Phi^2=0$ (and $M=0, q=2$).  (These points
reduce to the equilibrium set ${\cal K}_M$ in the absence of a magnetic field
consisting of Jacobs type I solutions with no matter and a massless scalar
field - Billyard et al., 1995). The eigenvalues are $\{0, 0, 2(1 - 2
\Sigma_+),
3(2-\gamma), 2(1+\Sigma_+\pm \sqrt{3} \Sigma_-), 6(1+ \frac{k
\Phi}{\sqrt{6}})\}$. There are two zero eigenvalues (corresponding to the
fact that ${\cal K}$ is a $2$-parameter family of equilibrium points). An
analysis shows that there is always a subset of ${\cal K}$ that acts as sources
(these are, in fact, global sources);
 all others are saddles.
\end{enumerate}

This contrasts with the situation in the absence of a scalar field in which
LeBlanc et al. (1995) showed that there are no equilibrium points that are
sources and argue that generically an orbit with $\Omega > 0$ will pass through
a transient stage and then approach  the Kasner circle into the past
(consisting of Kasner, Bianchi type I vacuum models).  Since these equilibrium
points are saddles, the orbit subsequently leaves along a uniquely determined
heteroclinic orbit and then approaches the Kasner circle again.  This process
then continues indefinitely, and the orbit follows an infinite heteroclinic
sequence of Rosen orbits and Taub orbits joining Kasner equilibrium points
undergoing Mixmaster-like chaotic oscillations (the past attractor here
consists of the union of the Kasner circle and a family of Rosen magneto-vacuum
type I orbits and Taub vacuum type II orbits -- see also Weaver, 2000).

\section{Bianchi I models}

The evolution equations for the Bianchi I models are obtained by restricting
the above equations to the zero-curvature case (i.e., $N_{\pm} = 0$), whence we
obtain
\be
 X= \left( \Sigma_+, \Sigma_-, \Phi, \Psi, \Omega\right)\in {\mathbb{R}}^5,
\ee
and the evolution equations are as follows:
\ba
\Sigma_+'\li=\li (q - 2)\Sigma _{+} + 2(1 - \Phi ^{2} - \Psi -
\Sigma_+^{2} -
\Sigma_{-}^2 - \Omega),\\
\Sigma_-'\li=\li (q - 2)\Sigma_{-},\\
\Phi'\li =\li (q - 2)\Phi  -\frac{\sqrt{6}k}{2}\Psi,\\
\Psi'    \li =\li (2q+2  + \sqrt{6}k\Phi)\Psi,\\
\Omega' \li  =\li (2q - 3\gamma  + 2)\Omega,
\ea
where a prime denotes differentiation with respect to the time $\tau$ and the
deceleration parameter $q$ is now given by
\be
q  = 1 + \Sigma_+^{2} + \Sigma_{-}^{2} + \frac{1}{2}(3\gamma  - 4)\Omega +
\Phi
^{2} - 2\Psi.
\ee
The magnetic field satisfies
\be
\frac{3}{2} M^2 = 1 - \Phi ^{2} - \Psi - \Sigma_+^{2} - \Sigma_{-}^2 -
\Omega,
\ee
and since $M^2\ge 0$, all of the variables are bounded: $0 \le \left\{
\Sigma_+^{2}, \Sigma_{-}^{2}, \Phi ^{2},\Psi, \Omega\right\} \le 1$.

The equations (3.1)-(3.6) can also be easily derived from the metric directly
in coordinate form -- see the Appendix.

\subsection{Invariant sets and Monotonic functions}

The existence of monotonic functions rule out periodic and recurrent orbits in
the phase space and enables global results to be obtained. When $\Omega
\ne 0$,
we can define $Z \equiv \Omega \Sigma_{-}^{-2}$, whence from the evolution
equations above we find that
\be
Z' = 3(2 - \gamma) Z,
\ee
so that $Z$ is a monotonically increasing function for $\gamma<2$ (LeBlanc et
al., 1995). From the monotonicity principle (Wainwright and Ellis, 1997) we can
deduce that $\Omega \rightarrow 0$ at early times and
$\Sigma_{-}^{2}\rightarrow 0$ at late times.

When $\Omega = 0$, we note that
\be
q - 2  = 3\left(M^2 + \Sigma_+^{2} + \Sigma_{-}^{2}  + \Phi ^{2} - 1\right) \le
0,
\ee
and so from equation (3.3) the variable $\Sigma_{-}$ is itself a monotonic
function.

There are a number of invariant sets. $M=0$ is an invariant set corresponding
to the absence of a magnetic field. The dynamics in this invariant set was
studied in Billyard et al. (1999). $\Phi = \Psi =0$ is an invariant set
corresponding to the absence of a scalar field. The dynamics in this invariant
set was studied in Collins (1972). $\Omega =0$ is an invariant set
corresponding to the absence of barotropic matter. This invariant set is
important for studying the the early time asymptotic behaviour. Finally,
$\Sigma_{-} =0$ is an invariant set which is important in the study of late
time asymptotic behaviour.

\subsection{Stability of Equilibria}

Let us present all of the equilibrium points and their associated eigenvalues.
We shall assume here that $1 \leq \gamma <2$.  The case of the bifurcation
value $\gamma =2$ will be discussed in Quinlan (2002). All equilibrium points
correspond to self-similar cosmological models.

Equilibrium Points: $\{\Sigma_+,\Sigma_-,\Phi,\Psi,\Omega\}$:

\subsubsection{Zero magnetic field}
\begin{enumerate}
\item[$\bullet$] ${\cal P}$ - Power-law inflation
  (flat FLRW):
$$\brac{0,0,-\;{k}{\sqrt{6}},1-\;{k^2}{6},0},~~M=0,~~q=\:12k^2-1.$$\\
        Eigenvalues:
 $\bras{k^2-4,k^2-3\gamma,\:12k^2-3,\:12k^2-3,\:12k^2-3}$ \\
        Exists provided $k^2\leq 6$.
        Sink if $k^2<4$ and $k^2<3\gamma$; saddle otherwise. Inflationary
if $k^ 2<2$
        (note that this is the {\it only} equilibrium point that can be
inflationary).

\item[$\bullet$] ${\cal F}_S$ - Matter-scaling (flat FLRW; Copeland et
al.,
  1998):
        $$\brac{0,0,-\;{\sqrt6}{2}\;\gamma
k,\;32\;{\gamma}{k^2}\bra{2-\gamma},
        \;{k^2-3\gamma}{k^2}},~~M=0,~~q=\:32\gamma-1>0.$$ \\
        Eigenvalues: $ \bras{\:32(\gamma-2),\:32(\gamma-2),3\gamma-4,
        \:34\bra{\gamma-2}\pm\;{3}{4k}\sqrt{\bra{2-\gamma}\bra{24\gamma^2
        +k^2\bras{2-9\gamma}}}}$ \\
        Exists provided  $k^2>3\gamma$. Sink  if $\gamma<\:43$; saddle
otherwise.

  \item[$\bullet$] ${\cal F}$ - Einstein-de-Sitter (flat FLRW):
        $$\{0,0,0,0,1\}~~M=0,~~q=\:32\gamma-1>0.$$\\
    Eigenvalues:  $\bras{\:32\bra{\gamma-2},\:32\bra{\gamma-2},
        \:32\bra{\gamma-2},3\gamma,3\gamma-4}$ \\
     Saddle.

  \item[$\bullet$] ${\cal K}$ - Kasner vacuum/massless scalar field:

$$\brac{\Sigma_+^2+\Sigma_-^2+\Phi^2=1,~\Psi=\Omega=0},~~M=0,~~q=2.$$\\
        Eigenvalues:
 $\bras{0,0,3(2-\gamma),2-4\Sigma_+,6+\sqrt{6}k\Phi}$ \\
        This is a 2-parameter family of equilibrium points. Sources when
$\Sigma _+<\;12$; saddles
        otherwise.

\end{enumerate}

\subsubsection{Non-zero magnetic field}
These all correspond to anisotropic models $(\Sigma^2_+ + \Sigma^2_- \neq 0)$.

\begin{enumerate}

  \item[$\bullet$] ${\cal J}$ - Matter and no scalar field
  (Jacobs magnetic field model; Jacobs, 1969):
        $$\brac{\:14(3\gamma-4),0,0,0,\:32\bra{1-\:14\gamma}},
        ~~M^2=\:18\bra{2-\gamma}\bra{3\gamma-4},~~q=\:32\gamma-1>0.$$

        Eigenvalues:
 $\bras{\:32\bra{\gamma-2},\:32\bra{\gamma-2},3\gamma,
        \:34\brac{\bra{\gamma-2}\pm\sqrt{\bra{2-\gamma}
        \bra{3\gamma^2-17\gamma+18}}}}$ \\
        Exists provided $\gamma> \:43$ (same as ${\cal F}$ for
$\gamma=\:43$). Saddle.

  \item[$\bullet$] ${\cal S}$ - Matter and scalar field:
        $$\brac{\:14(3\gamma-4),0,-\;{\sqrt6}{2}\;\gamma
        k,\;32\;{\gamma}{k^2}\bra{2-\gamma},
        \;{3}{8k^2}\bra{4k^2-k^2\gamma-8\gamma}},~~
        M^2=\:18\bra{2-\gamma}\bra{3\gamma-4},~~q=\:32\gamma-1>0.$$

Eigenvalues:
$\bras{\;32(\gamma-2),\;38\brac{2\bra{\gamma-2}\pm\;1k\sqrt{2\bra{2-\gamma}\bra
{a\pm b}}}
        }$

        where
        $
        a = 24\gamma^2+3\gamma^2 k^2-26k^2\gamma+20k^2$, and
$b =
\sqrt{\bra{18k^2\gamma-4k^2+a-32k\gamma}\bra{18k^2\gamma-4k^2+a+32k\gamma}}.$

Exists provided $\gamma>\:43$ (same as ${\cal F_S}$ for $\gamma=\:43$) and
$\gamma\leq\;{4k^2}{k^2+8}$. Sink.

  \item[$\bullet$] ${\cal V}$ - Scalar field and no matter:
 $$\brac{2\;{k^2-4}{k^2+8},0,-\;{2\sqrt{6}
k}{k^2+8},12\;{8-k^2}{\bra{k^2+8}^2},0},
~M^2=2\;{\bra{k^2-4}\bra{8-k^2}}{\bra{k^2+8}},~~q=\;{5k^2-8}{k^2+8}>0.$$

Eigenvalues:  $\bras{3\;{k^2-8}{k^2+8},3\;{k^2-8}{k^2+8},
        3\;{4k^2-k^2\gamma-8\gamma}{k^2+8},\;32\brac{\;{k^2-8}{k^2+8}
        \pm\sqrt{\;{17k^2-72}{k^2-8}}}}.$

Exists provided  $4\leq k^2\leq 8$. Sink when $\gamma>\;{4k^2}{k^2+8}$; saddle
otherwise.
\end{enumerate}

\subsection{Discussion}

\begin{figure}
\centerline{\psfig{file={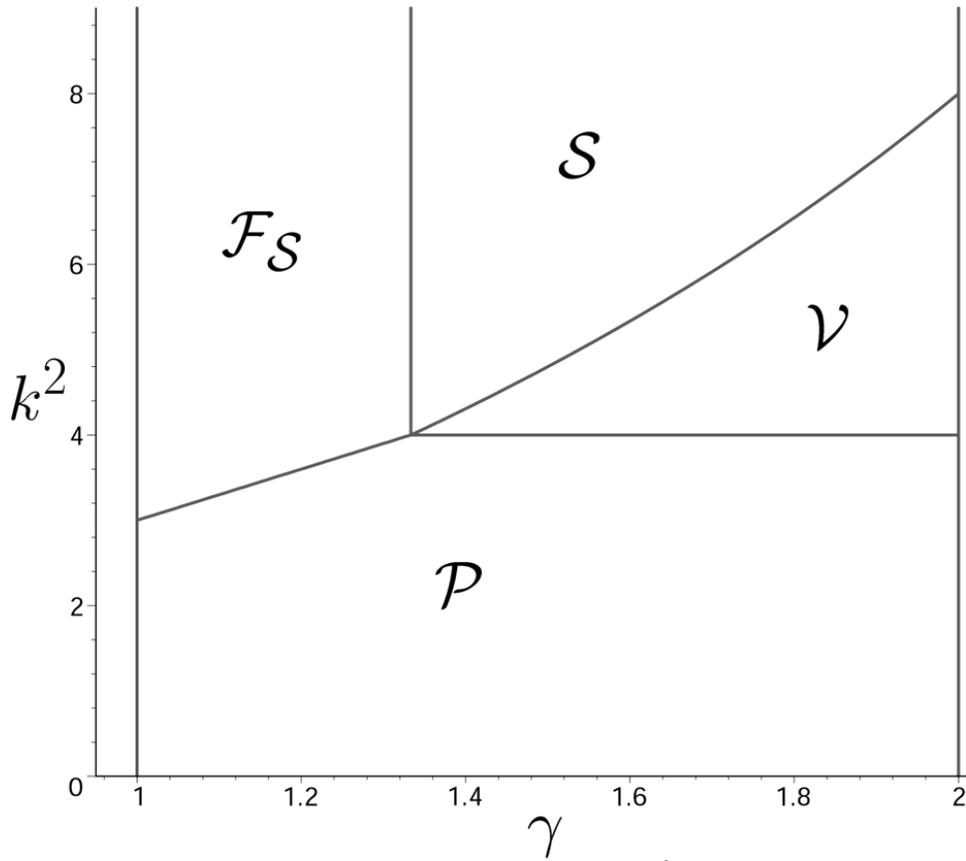},width=5in,angle=0}}
\caption{Summary of all of the sinks for the Bianchi I magnetic field
models in $\{\gamma, k^2\}$-parameter space. Note that for each specific pair
$\{\gamma, k^2\}$, there is a unique sink.\label{fig1}}
\end{figure}

In Table~\ref{table} we summarize the equilibrium points and their stability
for different values of the parameters $(\gamma,k^2)$. In Figure~\ref{fig1} we
show the sinks in $(\gamma,k^2)$-parameter space. The equilibrium point  ${\cal
P}$ represents the power-law FLRW solution with no matter and no magnetic
field, and is inflationary for $k^2<2$. It is a global attractor for
$k^2<3\gamma$ in the Bianchi I invariant set. Note that ${\cal P}$ is a sink in
the full class of Bianchi VI$_{\mbox{o}}$ models, unlike all other sinks in
Figure 1 which are only attractors in the Bianchi I invariant set. Also note
that the monotonic function $Z$ in the Bianchi I invariant set indicates that
$\Omega\rightarrow0$ at early times (at the sources) and $\Sigma_-\rightarrow0$
at late times (at the sinks), consistent with the results in Table~\ref{table}.

The equilibrium points  ${\cal S}$ and  ${\cal V}$, in which all of the matter
fields are non vanishing and in which matter field (only) vanishes,
respectively, correspond to new exact self-similar magnetic field cosmological
models. The new solutions are given explicitly in the Appendix.

\begin{figure}
\centerline{\psfig{file={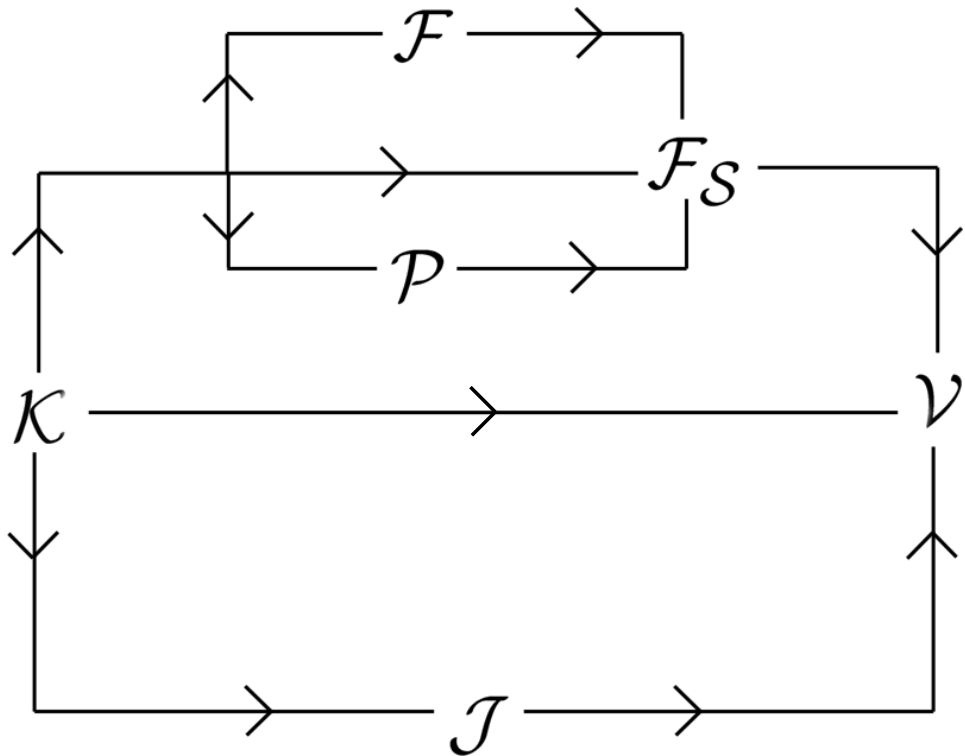},width=5in,angle=0}}
\caption{Heteroclinic sequences for the Bianchi I models when
$\gamma = \frac{5}{3}$, $5 < k^2 < \frac{40}{7}$.  \label{fig2}   }
\end{figure}

Heteroclinic sequences (and hence the transient behaviour of the Bianchi I
models) can easily be deduced from the Table. As a particular example, for
$\gamma=5/3$ and $5<k^2<40/7$, a subset of $\cal K$ constitute sources (the
remainder are saddles), $\cal V$ is a sink, $\cal J$ is a saddle (that lies in
the $\Phi=0,\Psi=0$ ($M=0$) submanifold) and $\cal F$, $\cal P$ and ${\cal
F}_S$ are saddles (that lie in the isotropic $\Sigma_+ =0$, $\enskip
\Sigma_-=0$ ($M=0$) submanifold--in this submanifold $\cal K$ is a source,
$\cal F$ and $\cal P$ are saddles, and ${\cal F}_S$ is a sink). The
heteroclinic sequence in this case is presented in Figure~\ref{fig2}.

From the the Table and Figures we can deduce both the asymptotic and the
intermediate behaviour of the Bianchi I models. Note that as $\gamma$
increases, the magnetic field becomes increasingly important at late times,
consistent with the observation of Wainwright (2000) (in the absence of a
scalar field).

\begin{table}[ht](a)~~$1\leq\gamma\leq\frac{4}{3}$
\begin{tabular}{|c|ccccccc}
  $k^2$        &&   $3\gamma$ & $4$ & $6$ & $8$ \\
  \hline\hline
  ${\cal P}$   &  {\vphantom{\Large X}\textbf{SINK}}&\vline &
            {\textrm{SADDLE}}&\vline & ~~~~~~~{\scriptsize\textsc{does not exist}}
            \\ \hline
  ${\cal F_S}$ &{\scriptsize\textsc{does not exist}}& \vline &
            &\multicolumn{2}{c}{\vphantom{\Large X}\textbf{SINK}}&  \\ \hline
  ${\cal F}$   & \multicolumn{6}{c}{\vphantom{\Large X}\textrm{SADDLE}} \\ \hline
  ${\cal V}$   &&{\scriptsize\textsc{does not exist}}&
            \vline&{\vphantom{\Large X}\textrm{SADDLE}}&\vline
            & {\scriptsize\textsc{does not exist}}~~~~~~~ \\
\end{tabular}
\vskip 5mm
(b)~~$\frac{4}{3}<\gamma<2$
\begin{tabular}{|c|ccccccccccc}
  $k^2$        &&4&   $3\gamma$ & 6 & $8\gamma/(4-\gamma) $ && $8$ &\hphantom{xxx}\\
  \hline\hline
  ${\cal P}$   &  {\vphantom{\Large X}\textbf{SINK}}&\vline &
            {\textrm{SADDLE}}&\vline &&{\scriptsize\textsc{does not exist}} \\ \hline
  ${\cal F_S}$ &{\scriptsize\textsc{does not exist}}&& \vline
            &\multicolumn{3}{c}{\vphantom{\Large X}\textrm{SADDLE}}&  \\ \hline
  ${\cal F}$   & \multicolumn{6}{c}{\vphantom{\Large X}\textrm{SADDLE}} \\ \hline
  ${\cal J}$ & \multicolumn{6}{c}{\vphantom{\Large X}\textrm{SADDLE}} \\ \hline
  ${\cal S}$ &&&{\vphantom{\Large X}\scriptsize\textsc{does not exist}}&&\vline &
            {}\textbf{SINK} \\ \hline
  ${\cal V}$ &{\scriptsize\textsc{does not exist}}&\vline&
            ~~~~{\vphantom{\Large X}\textbf{SINK}}&&\vline&
            {\vphantom{\Large X}\textrm{SADDLE}}~~~~&\vline&
            {\scriptsize\textsc{does not exist}} \\
\end{tabular}
\caption{Summary of all of the equilibrium points (as defined in
the text) and their nature for different $\{\gamma, k^2\}$-parameter values in
the Bianchi I models with (a) $1 \leq \gamma \leq \frac{4}{3}$ and (b) $\gamma
>\frac{4}{3}$.  Note that for $\frac{4}{3} < \gamma < 2$, ~$4 < 3 \gamma < 8
\gamma(4-\gamma)^{-1} < 8$, and in (b) we have identified
$8\gamma(4-\gamma)^{-1} >6$ (i.e., $\gamma > \frac{12}{7}$) to be specific. For
all values of $\gamma$,  a subset of ${\cal K}$ are sources (the remainder are
saddles).\label{table}}
\end{table}

\section{Conclusions}

We  have investigated the class of open Bianchi  VI$_{\mbox{o}}$ universe
models with barotropic matter, a scalar field with an exponential potential and
a uniform magnetic field. We discussed some of the general properties of the
models by utilyzing dynamical systems techniques. The equilibrium point ${\cal
P}$ (with $M =0$), corresponding to the power-law inflationary flat FLRW model,
is a global sink for $k^2<2$. Hence all models are future aymptotic to this
inflationary attractor for these parameter values. There is an equilibrium set
${\cal K}$ with $1-\Sigma^2_+ - \Sigma^2_- - \Phi^2=0$ (and $M=0, q=2$), a
subset of which are global sources. Hence all models are past asymptotic to
massless scalar field models with no matter and no magnetic field.

A (partial) analysis of the saddles was undertaken in order to determine some
of the transient features of the models. In particular, we found that there are
no equilibrium points with $M \neq 0$ in the Bianchi VI$_{\mbox{o}}$ state
space which are inflationary, and hence concluded that inflation is not
fundamentally affected by the presence of a uniform primordial magnetic field
in these models.

This latter result is not necessarily inconsistent with the conclusions of
Matravers and Tsagas (2000), since only a uniform magnetic field was considered
here. Indeed, it is the non-uniform magnetic field gradients that give rise to
the magneto-curvature effects in their work which modify the cosmological
expansion rate of an almost-FLRW universe and that can have undesirable
implications for inflationary models (and may even prevent inflation taking
place in the presence of primordial magnetism). It might be thought that on
large scales that the magnetic field will be approximately homogeneous, but
Matravers and Tsagas (2000) have argued that even weak magetic effects may have
significant cosmological consequences. The drawback of their work is that only
a local perturbation analysis was performed and questions of genericity and
long term behaviour cannot be easily addressed. Clearly, an investigation of
the qualitative properties of a class of scalar field cosmological models with
an inhomogeneous magnetic field would extend and generalize both the work of
Matravers and Tsagas (2000) and the (restricted) analysis of a uniform magnetic
field here.

In order to investigate the intermediate behaviour of the models, and hence
their physical properties, we discussed the (subset) of Bianchi I models in
more detail. We found the equilibrium points  ${\cal S}$ and  ${\cal V}$, in
which all of the matter fields are non vanishing and in which the matter field
(only) vanishes, respectively (corresponding to new exact self-similar magnetic
field cosmological models), act as sinks in the Bianchi I subset. Heteroclinic
sequences (and hence the transient behaviour of the Bianchi I models) were
discussed. We found that as $\gamma$ increases, the magnetic field becomes
increasingly important at late times.

\acknowledgements

This work has been supported, in part, by the Natural Sciences and Engineering
Research Council of Canada. We would like to thank Christos Tsagas for helpful
comments.

\appendix

\section{}

In the main text we followed the dynamical systems theory approach of LeBlanc
et al. (1995) to  analyse the evolution of orthogonal Bianchi cosmologies of
type VI$_{\mbox{o}}$ with a uniform magnetic field. This approach uses an
invariant orthonormal frame  in which the commutation functions are  the basic
variables. Expansion-normalized variables are then introduced, and a reduced
set of evolution equations (the dynamical system) for the dimensionless shear
variables and spatial curvature variables  as well as the dimensionless
magnetic field variables and the density parameter are then obtained. All of
the equations in the text follow from the analysis of LeBlanc et al. (1995),
wherein all quantities are properly defined. However, a coordinate approach is
also possible, and it may be of use for the reader to get a better sense of how
the equations are derived, at least in the case of the Bianchi I models.

The metric for the class of anisotropic Bianchi I models, which are the
simplest spatially homogeneous generalizations of the flat FLRW models which
have non-zero shear but zero three-curvature, is given by
\be
ds^2 = -dt^2 + X^{2}dx^2 + Y^{2}dy^2 + Z^{2}dz^2,
\ee
where $X,Y,Z$ are functions of $t$ only. The expansion rate is given by
\be
\theta \equiv 3H = \frac{\dot{X}}{X}+ \frac{\dot{Y}}{Y}+ \frac{\dot{Z}}{Z},
\ee
and when $Y=Z$ (the LRS subcase) there is only one independent rate of shear,
which is given by
\be
\sigma = \frac{1}{3}\left(\frac{\dot{Y}}{Y} - \frac{\dot{X}}{X}\right).
\ee
Normalized (dimensionless) variables and a logarithmic time variable $\tau$ are
defined by equations (2.3) and (2.4), which then lead to the evolution
equations (3.1)-(3.6) in coordinate form (Collins, 1972).

\subsection{Exact Solutions}

We can write down the exact solutions corresponding to the equilibrium points
${\cal S}$ and ${\cal V}$ in the text.

\begin{itemize}

\item ${\cal S}$:  Since $q=\:32\gamma-1$, we can immediately integrate
the Raychaudhuri equation  $H'/H = -(1+ q)$ to obtain $H = \frac{2}{3\gamma}
t^{-1}$. Since $\Sigma_{-} = 0$ and  $\Sigma_{+} =\:14(3\gamma-4)$, it follows
that $Y=Z$,
 so that $\sigma = \Sigma_{+} \times H  = \frac{1}{2}(1-\frac{4}{3\gamma})t^{-1}$.
The metric is consequently of power-law form, viz.,
\be
ds^2 = -dt^2 + t^{2 p_1}dx^2 + t^{2 p_2}(dy^2 + dz^2),
\ee
where $p_1$ and $p_2$ are constants. From the expressions for $H$ and   $\sigma
$ we then find that $p_1 + 2 p_2 = \frac{2}{\gamma}$ and $p_2 - p_1 =
\frac{3}{2}  -
\frac{2}{\gamma}$ and hence
\be
p_1  =  \frac{2}{\gamma} - 1, ~~~p_2 = \frac{1}{2}.
\ee
The geometry is that of the Jacobs magnetic field model (Jacobs, 1969), and the
only difference here is that the matter fields constituting the source are
given by
\be
\rho \equiv 3\Omega H^2 =\bra{ \frac{4k^2 - k^2 \gamma -8 \gamma}{2 k^2 \gamma^2}}t^{-2},
~~~~H_1 \equiv 3 \sqrt{2} M H =
\frac{\sqrt{(2-\gamma)(3\gamma-4)}}{\gamma}t^{-1},
\ee
and the scalar field is given by
\be
\phi = \phi_0 -  \frac{2}{k} \ln(t),
\ee
where $ \phi_0 = \frac{1}{k} \ln\bra{\frac{2(2-\gamma)}{\gamma k^2 V_0}}$. This
solution exists provided $\gamma>\:43$.

\item  ${\cal V}$: In this solution $q=\;{5k^2-8}{k^2+8}$, so that $H =
\frac{k^2+8}{6k^2} t^{-1}$. $\Sigma_{-} = 0$ and $\Sigma_{+} = 2
\;{k^2-4}{k^2+8}$, and so the metric is of the form of equation (A4), where now
\be
p_1 =  \frac{8-k^2}{2 k^2}, ~~~p_2 = \frac{1}{2}.
\ee
There is no perfect fluid in this solution ($\rho =0$), and the magnetic field
is given by
\be
H_1 = \frac{\sqrt{\bra{k^2-4}\bra{8-k^2}\bra{k^2+8}}}{k}t^{-1}.
\ee
The scalar field is given by equation (A7) with $ \phi_0 = \frac{1}{k}
\ln\bra{\frac{8-k^2}{k^4 V_0}}$. This new solution exists provided  $4\leq k^2\leq
8$.

\end{itemize}

\end{document}